\documentclass[superscriptaddress,aps,preprintnumbers,amsmath,showpacs,amssymb,prd,nofootinbib,reprint]{revtex4-1}
\usepackage{bm, color} 
\usepackage{amssymb,amsfonts,slashed,amsthm,amsmath,graphicx, soul}
\usepackage{epsfig}
\usepackage{ulem}
\usepackage{xcolor}

\begin{document}


\renewcommand{\figurename}{Fig.}
\renewcommand{\tablename}{Table.}
\newcommand{\Slash}[1]{{\ooalign{\hfil#1\hfil\crcr\raise.167ex\hbox{/}}}}
\newcommand{\bra}[1]{ \langle {#1} | }
\newcommand{\ket}[1]{ | {#1} \rangle }
\newcommand{\beq}{\begin{equation}}  \newcommand{\eeq}{\end{equation}}
\newcommand{\bef}{\begin{figure}}  \newcommand{\eef}{\end{figure}}
\newcommand{\bec}{\begin{center}}  \newcommand{\eec}{\end{center}}
\newcommand{\non}{\nonumber}  \newcommand{\eqn}[1]{\begin{equation} {#1}\end{equation}}
\newcommand{\laq}[1]{\label{eq:#1}}  
\newcommand{\dd}[1]{{d \o d{#1}}}
\newcommand{\Eq}[1]{Eq.~(\ref{eq:#1})}
\newcommand{\Eqs}[1]{Eqs.~(\ref{eq:#1})}
\newcommand{\eq}[1]{(\ref{eq:#1})}
\newcommand{\Sec}[1]{Sec.\ref{chap:#1}}
\newcommand{\ab}[1]{\left|{#1}\right|}
\newcommand{\vev}[1]{ \left\langle {#1} \right\rangle }
\newcommand{\bs}[1]{ {\boldsymbol {#1}} }
\newcommand{\lac}[1]{\label{chap:#1}}
\newcommand{\SU}[1]{{\rm SU{#1} } }
\newcommand{\SO}[1]{{\rm SO{#1}} }
\def\({\left(}
\def\){\right)}
\def\dt{{d \o dt}}
\def\diag{\mathop{\rm diag}\nolimits}
\def\Spin{\mathop{\rm Spin}}
\def\O{\mathcal{O}}
\def\U{\mathop{\rm U}}
\def\Sp{\mathop{\rm Sp}}
\def\SL{\mathop{\rm SL}}
\def\tr{\mathop{\rm tr}}
\def\ebq{\end{equation} \begin{equation}}
\newcommand{\OR}{~{\rm or}~}
\newcommand{\AND}{~{\rm and}~}
\newcommand{\EV}{ {\rm \, eV} }
\newcommand{\KEV}{ {\rm \, keV} }
\newcommand{\MEV}{ {\rm \, MeV} }
\newcommand{\GEV}{ {\rm \, GeV} }
\newcommand{\TEV}{ {\rm \, TeV} }
\def\o{\over}
\def\a{\alpha}
\def\b{\beta}
\def\c{\varepsilon}
\def\d{\delta}
\def\e{\epsilon}
\def\f{\phi}
\def\g{\gamma}
\def\h{\theta}
\def\k{\kappa}
\def\l{\lambda}
\def\m{\mu}
\def\n{\nu}
\def\p{\psi}
\def\q{\partial}
\def\r{\rho}
\def\s{\sigma}
\def\t{\tau}
\def\u{\upsilon}
\def\w{\omega}
\def\x{\xi}
\def\y{\eta}
\def\z{\zeta}
\def\D{\Delta}
\def\G{\Gamma}
\def\H{\Theta}
\def\L{\Lambda}
\def\F{\Phi}
\def\P{\Psi}
\def\S{\Sigma}
\def\me{\mathrm e}
\def\ol{\overline}
\def\tl{\tilde}
\def\*{\dagger}


\preprint{TU-1174}

\title{
Stability of domain wall network with initial inflationary fluctuations and its implications for cosmic birefringence
}

\author{
Diego Gonzalez
}

\affiliation{Department of Physics, Tohoku University, 
Sendai, Miyagi 980-8578, Japan}

\author{
Naoya Kitajima
}
\affiliation{Frontier Research Institute for Interdisciplinary Sciences, Tohoku University, Sendai, Miyagi 980-8578, Japan}
\affiliation{Department of Physics, Tohoku University, 
Sendai, Miyagi 980-8578, Japan}

\author{
Fuminobu Takahashi
}
\affiliation{Department of Physics, Tohoku University, 
Sendai, Miyagi 980-8578, Japan}

\author{
Wen Yin
}
\affiliation{Department of Physics, Tohoku University, 
Sendai, Miyagi 980-8578, Japan}

\begin{abstract}
We study the formation and evolution of domain walls with initial inflationary fluctuations by numerical lattice calculations {that, for the first time, correctly take} into account correlations on superhorizon scales. 
We find that, contrary to the widely-held claim {over the past few tens of years},  the domain wall network exhibits remarkable stability even when the initial distribution is largely biased toward one of the minima. This is due to the fact that
the domain wall network retains information about initial conditions on superhorizon scales, and that the scaling solution is not a local attractor in this sense. {Our finding immediately implies that such domain walls will have a significant impact on cosmology, including the production of gravitational waves, baryogenesis, and dark matter from domain walls.
}
Applying this result to the axion-like particle domain wall, we show that it not only explains the isotropic cosmic birefringence suggested by the recent analysis, but also predicts anisotropic cosmic birefringence that is nearly scale-invariant on large scales and can be probed by future CMB observations.

\end{abstract}

\maketitle
\flushbottom

\vspace{1cm}

\textit{\textbf{Introduction.\,---\,}} 
Topological defects are formed by spontaneous symmetry breaking and have been widely studied.
~\cite{Zeldovich:1974uw,Kibble:1976sj,Vilenkin:1981zs} (see also e.g.~\cite{Vilenkin:2000jqa} for a review).
While the stability of individual topological defects is guaranteed by the conservation of topological charge, the evolution of the large number of topological defects formed in the early universe is not obvious because it depends on their evolution and interactions.
Such topological defects and their traces may be explored experimentally in the present universe, or in some cases they may pose serious cosmological problems.

We focus on the cosmological evolution of the simplest class of topological defects, domain walls (DWs). DWs are field configurations connecting distinct degenerate vacua and are formed from spontaneous breaking of discrete symmetries. The simplest DWs are the ones associated with 
the $Z_2$ symmetric potential,
\begin{equation}
V(\phi) = \frac{\l}{4} (\f^2-v^2)^2
= \frac{\l}{4} v^4 - \frac{1}{2} m_0^2 \phi^2 + \frac{\lambda}{4} \phi^4
\label{pot}
\end{equation}
with $m_0 \equiv \sqrt{\lambda} v$. 
Here $\phi$ is the scalar field, $\l (> 0)$ is the quartic coupling, and  two degenerate vacua are located at $\phi_{\rm min}=\pm v$. Once such DWs separated by the two vacua are formed in the early universe, it is known that the subsequent evolution converges to the so-called scaling solution, a situation where there is on average about one DW per the Hubble horizon~\cite{Press:1989yh,Hindmarsh:1996xv,Garagounis:2002kt,Oliveira:2004he,Avelino:2005kn,Leite:2011sc,Leite:2012vn,Martins:2016ois}.

The energy density of DWs that obey the scaling solution decreases more slowly than radiation or matter, and thus easily dominates the universe, making it significantly inhomogeneous. Even if it were subdominant, it could be inconsistent with current observations by making the cosmic microwave background (CMB) spectrum largely anisotropic.  This is the notorious cosmological DW problem~\cite{Zeldovich:1974uw, Vilenkin:1984ib}.  This problem can be avoided if the DW tension is small enough, or if DWs are unstable. A well-known way to destabilize DWs is to introduce 
a symmetry-breaking bias in the potential
and/or the initial distribution of the scalar field~\cite{Vilenkin:1981zs,Sikivie:1982qv,Mohanty:1984pj,Gelmini:1988sf,Lalak:1993ay,Lalak:1993bp,Lalak:1994qt,Coulson:1995nv,Coulson:1995uq,Larsson:1996sp,Correia:2014kqa,Correia:2018tty,Krajewski:2021jje}.

Our focus is on the case in which the initial distribution is biased. Such a biased initial distribution can be realized when fluctuations are generated by non-thermal mechanisms, {for example, fluctuations generated during inflation}~\cite{Linde:1990yj,Lyth:1991ub,Nagasawa:1991zr}. 
{
It was argued in Refs.~\cite{Lalak:1993ay,Lalak:1993bp,Lalak:1994qt,Coulson:1995uq,Coulson:1995nv,Larsson:1996sp} that since the DW is a nonlinear object and the field distribution is highly non-Gaussian, the DW network with initial inflationary fluctuations can be understood based on percolation theory. 
}
{Under this assumption}, previous numerical calculations showed that DWs decay quickly even when the bias in the initial distributions is very small~\cite{Lalak:1993bp,Lalak:1994qt,Coulson:1995uq,Coulson:1995nv,Larsson:1996sp,Correia:2014kqa,Correia:2018tty}. In this Letter we focus on the DWs generated by inflationary fluctuations and show that, contrary to this widely held conclusion, the DWs are very stable against biased initial distributions.

The purpose of this Letter is two-fold.
First, we study the formation and evolution of DWs with the initial inflationary fluctuations
by numerical lattice simulations, correctly taking into account correlations on superhorizon scales. We then show that
the resultant DW networks are very stable against biased initial fluctuations. This stability is due to the DW structure on large scales, which reflects the initial fluctuations at scales beyond the Hubble horizon. 

Second, as a phenomenological application of this result, we consider axion-like particle (ALP) DW with inflationary fluctuations.  In Ref.~\cite{Takahashi:2020tqv} two of the present authors (FT and WY)  proposed the ALP DW   to explain a hint for the isotropic cosmic birefringence (CB)~\cite{Minami:2020odp, Diego-Palazuelos:2022dsq},
$
\beta=0.36\pm 0.11 {\rm\, deg},
$
where $\beta$ denotes the isotropic rotation angle of the CMB polarization. Interestingly,
the ALP DWs naturally account for the closeness of the rotation angle to the fine structure constant,  $\alpha = 1/137 {\rm\, rad} \simeq 0.42 {\rm\, deg}$, over a very wide range of the ALP mass and decay constant. Furthermore, 
they predict a characteristic anisotropic CB determined by the DW distribution on the last scattering surface (LSS), but the case with the initial correlation on superhorizon scales has not been studied.
In this Letter, we investigate  for the first time the anisotropic CB power spectrum predicted by ALP DWs with inflationary fluctuations
and show that it can be tested in future CMB observations such as LiteBIRD~\cite{Matsumura:2016sri}, Simons Observatory~\cite{Ade:2018sbj}, CMB-S4-like~\cite{Abazajian:2016yjj}, and PICO~\cite{Hanany:2019lle}.

Here we mention some relevant past literature.
The axion DW with inflationary fluctuations was first considered in \cite{Linde:1990yj,Lyth:1991ub} in the context of the QCD axion, and since then it has been
discussed 
in various context~\cite{Khlopov:2004sc,Davoudiasl:2006ax,Daido:2015gqa,Daido:2015bva,Kobayashi:2016qld,Liu:2020mru, Sakharov:2021dim,Takahashi:2020tqv,Kitajima:2022jzz}.
In the symmetry breaking scenario during inflation, superhorizon-scale fluctuations are generated over a certain range of scales, and their spatial distribution was studied in Ref.~\cite{Nagasawa:1991zr}.
See also Refs.~\cite{Kobayashi:2016qld,Takahashi:2020tqv,Kitajima:2022jzz} for discussions on the relation between the superhorizon fluctuations and the DW formation.
However, to the best of our knowledge, detailed numerical lattice simulations on the formation and evolution of DWs that properly incorporate long-wavelength correlations have not been performed.

\textit{\textbf{Setup.\,---\,}} 
Let us consider a scalar field $\phi$ with the quartic potential (\ref{pot}) as a simple example. 
The scalar mass at the potential minimum is  $m_\phi = \sqrt{2}m_0$.
A planar DW perpendicular to the $z$-axis is given by
$
\phi_{\rm DW}= v \tanh {(z m_\phi/2)}.
$
Thus the DW width is of $m_\f^{-1}$, and the tension $\sigma$ is
$4\sqrt{\l} v^3/3$. As long as we focus on the macroscopic behavior of DW network at scales much larger than the DW width, the following discussion is also applicable to other potential forms, such as the sine-Gordon potential, which will be discussed later.

The equation of motion for the minimally coupled $\phi$ in the flat Friedmann-Lema\^itre-Robertson-Walker universe with three spatial dimensions is
\beq
\frac{\partial^2 {\phi}({\bm x},t)}{\partial t^2} + 3H \frac{\partial {\phi({\bm x},t)}}{\partial t} - \frac{\partial_i^2 \phi({\bm x},t)}{a^2}
+\frac{\partial V}{\partial \phi}
= 0, \label{EQM}
\eeq
where $t$ is the cosmic time, ${\bm x}$
is the comoving coordinate, $a(t)$ is the scale factor, $H=(da/dt)/a$ is the Hubble parameter, and $\partial_i^2$ is the Laplace operator.
Here we neglect the gravitational self-interaction of DWs, which is justified
if the DW energy density is subdominant.

\textit{\textbf{Clarification of initial fluctuations.\,---\,}} 
The initial conditions of the scalar field are of essential importance for the formation and subsequent evolution of DW, which is one of the findings of this Letter. We assume that the scalar field initially fluctuates around $\phi_0$, where $\phi_0$ is the field value averaged over the spatial region of our interest. We can bias the initial fluctuations by making $\phi_0$ nonzero. For simplicity, let us first assume $\phi_0=0$ and consider fluctuations around the origin.

The scale-dependence of the fluctuations is encoded in the power spectrum $P(k)$ defined by
\begin{equation}
\label{two-point-func}
\langle \phi(\bm{k})\phi(\bm{k}')\rangle = (2\pi)^d\delta^{(d)}(\bm{k}+\bm{k}') P(k),
\end{equation}
where  $\phi(\bm{k}) = \int d^{d} x e^{- i {\bm k} \cdot {\bm x}} \phi({\bm x})$ is the Fourier transform, the angle brackets mean an ensemble average, and we introduced the spatial dimensions $d = 2$ or $3$ for later use. Here,
the rotational invariance is assumed to use $k=|\bm k|$ for the power spectrum.
Then,  the variance of fluctuations in real space can be expressed as
\begin{equation}\label{eq:Pk}
\langle \phi(\bm{x})^2 \rangle = \int \frac{d^d{\bm k}}{(2\pi)^d} P({k}) = \int d\ln k\, \mathcal{P}(k),
\end{equation}
where $\mathcal{P}(k)$ is the reduced power spectrum defined by $\mathcal{P}(k)=k^2P(k)/2\pi$ for $d=2$ and $\mathcal{P}(k)=k^3P(k)/2\pi^2$ for $d=3$.
We mainly consider inflationary fluctuations;
if the scalar is sufficiently light,
i.e. $m_\phi \ll H_{\rm inf}$, it acquires Gaussian fluctuations with 
almost scale-invariant 
power spectrum $\mathcal{P}(k) \simeq (H_\mathrm{inf}/2\pi)^2$ during inflation, where $H_{\rm inf}$ is the inflationary Hubble parameter.

The spatial average $\vev{\phi}(=\phi_0)$ is generically non-zero. In this case the initial distribution is biased, and it is obtained simply by shifting the above distribution by $\phi_0$. To quantify the bias, we define
the bias parameter as 
\beq 
 {b_{d}}\equiv \frac{\vev{\f}}{\sqrt{\vev{(\f-\vev{\phi})^2}}}.
\eeq 
For the inflationary fluctuations, we replace the denominator with $\sqrt{{\cal P}(k)}$ to avoid the logarithmic dependence on the lattice box size. 

For the numerical lattice calculations, the initial conditions for the inflationary fluctuations are prepared as follows. First, we generate scalar fluctuations in the momentum space so that they satisfy the scale-invariant power spectrum. Then, we  set the initial scalar field fluctuations in the real space by the inverse Fourier transform. On the other hand, for the Gaussian white noise, as in the case of thermal fluctuations, we randomly generate a scalar field value at each spatial point, using the Gaussian probability distribution.

\textit{\textbf{Scaling solution.\,---\,}} 
For the white noise or thermal initial conditions, the DW network quickly converges to the scaling solution, 
where there is on average about one DW per the Hubble
horizon~\cite{Press:1989yh,Hindmarsh:1996xv,Garagounis:2002kt,Oliveira:2004he,Avelino:2005kn,Leite:2011sc,Leite:2012vn,Martins:2016ois}, given by 
\begin{equation}
    \frac{L_{\rm DW}}{H^{1-d}} = {\cal O}(1)
    \label{ss}
\end{equation}
with $L_{\rm DW}$ being the average DW length (area) per Hubble horizon for $d=2(3)$.
We will see that, while {the evolution of the DW length is given by  Eq.~(\ref{ss})} for inflationary fluctuations, the spatial DW distribution is significantly different from that for the white noise initial condition.

\textit{\textbf{Stability of the DWs against the initial condition bias.\,---\,}} 
We solve numerically the equation of motion (\ref{EQM}) on a lattice in the comoving coordinate space with two different initial conditions: inflationary fluctuations and Gaussian white noise. In the case of the inflationary fluctuations, in particular, the lattice box size should be sufficiently large because of the correlations on large scales. To this end, we focus on DWs with their motion restricted in 2D space. We have performed numerical simulations in 3D space to confirm that  similar results are obtained both quantitatively and qualitatively.

We have set the box size to contain $200^2$  Hubble horizons at the initial conformal time $\tau_i=1/m_0$ on a lattice with a grid size of $4096 \times 4096$ and a periodic boundary condition. Here we assume the radiation dominated universe, and  define the conformal time $\tau = \sqrt{2t/m_0}$ for convenience. 
Then, we have solved the equation of motion (\ref{EQM}) using the second order leapfrog method with the two initial conditions, varying the bias $b_d$ of the initial distributions.

The snapshots of the scalar field at $\tau=10/m_0$ in the case of $b_d = 0$ are shown in Fig.~\ref{fig:snap}, where the DWs separate the two vacua, $\phi = \pm v$. Compared to the white noise case, there are very large regions in the case of inflationary fluctuations. Such a region much larger than the Hubble horizon is considered to be stable, at least until the horizon grows to its size, due to the causality.  Thus, in the case of scale-invariant fluctuations, the DW network is expected to be very stable because such a large region exists at any scale.

\begin{figure}[!t]
\begin{center}  
\includegraphics[width=4.2cm]{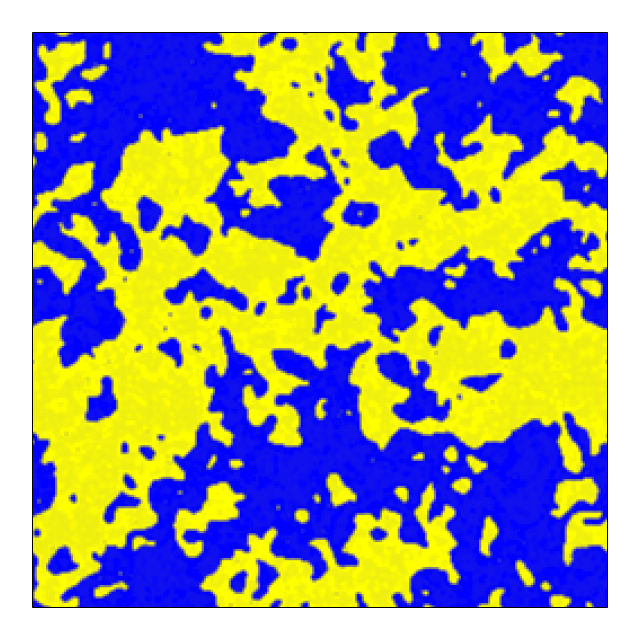}
\includegraphics[width=4.2cm]{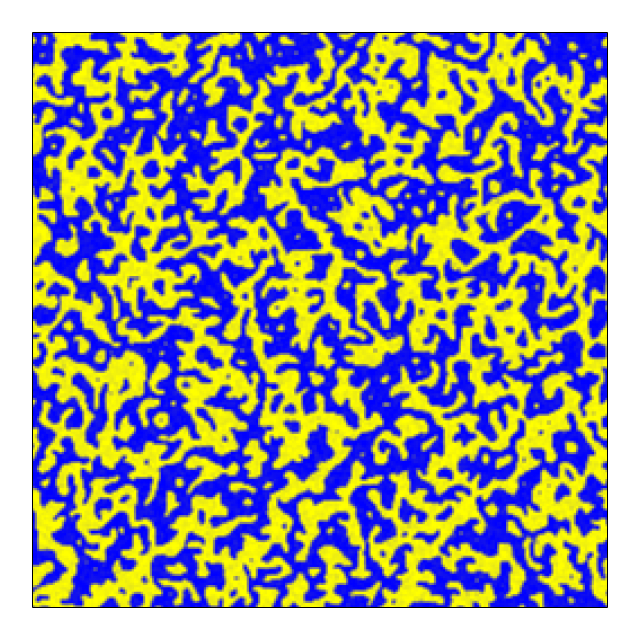}
    \end{center}
\caption{Snapshots of the DW network at $\tau = 10/m_0$ in the absence of the bias, $b_d = 0$. Blue and yellow regions correspond to $\phi = -v$ and $+v$, respectively. Initial fluctuations are set with  inflationary fluctuations (left) and Gaussian white noise (right). The box contains $\sim 40^2$ horizons. 
} \label{fig:snap}
\end{figure}
\begin{figure}[h]
\includegraphics[width=8cm]{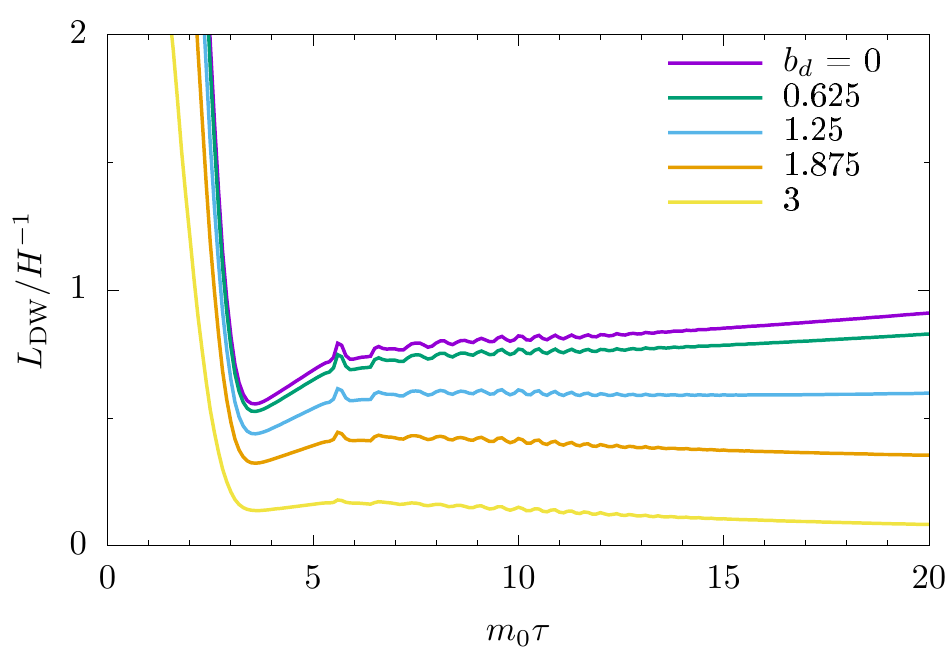}
\includegraphics[width=8cm]{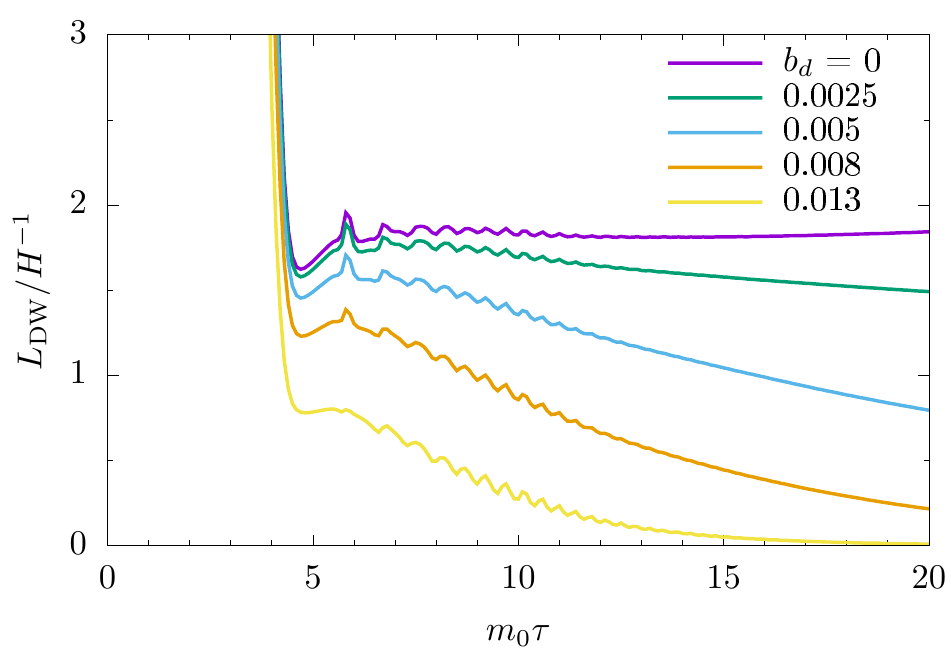}
\caption{The averaged DW length per Hubble horizon as a function of $\tau$, based on 500 simulations. Initial fluctuations are set with inflationary fluctuations (top) and Gaussian white noise (bottom).
The top line in each panel corresponds to  $b_d=0$, and the lower lines are for more biased initial fluctuations.} \label{fig:DWlength}
\end{figure}

To see the stability of DWs with biased initial fluctuations,
we show in Fig.~\ref{fig:DWlength} the time evolution of the averaged DW length per Hubble horizon.
In each panel the top line corresponds to $b_d = 0$ which satisfies Eq.~(\ref{ss})\footnote{The slight increase of $L_{\rm DW}/H^{-1}$ is mainly due to the finite box effect, which becomes milder as the lattice size increases.}, while the lower lines are for more biased cases. 
In the case of white noise, the DWs collapse in a very short timescale even with a very small bias $b_d \ll 1$, whereas in the case of inflationary fluctuations, DWs are very long-lived even with a very large bias  $b_d={\cal O}(1)$. The reason for the stability
is the correlation of the scalar fluctuations at large scales. This leads to the characteristic void structure of domains, which cannot 
collapse until the Hubble horizon becomes comparable to its size. This is the central result of this Letter, which differs significantly from the claims in the literature~\cite{Lalak:1993ay,Lalak:1993bp,Lalak:1994qt,Coulson:1995uq, Coulson:1995nv,Larsson:1996sp}.

Lastly, we show in Fig.~\ref{fig:spectrum} the reduced power spectrum defined by Eq.~(\ref{eq:Pk}) with $d=2$ at $ \tau = 10/m_0$ for the case of $b_d=0$ on a lattice with a grid size of $16384 \times 16384$ ($4096 \times 4096$) for the scale-invariant (white-noise) initial condition.
Here   $k_H$ is the comoving wavenumber such that its corresponding wavelength is equal to the Hubble horizon, $k_H/a = 2\pi H$. 
The red and blue lines correspond to the initial conditions of inflationary fluctuations and white noise~\cite{Kitajima:2022jzz}, respectively. Each light-colored band represents the $1 \sigma$ statistical uncertainties based on the $50$ times simulations. 
One can see that  the power spectrum is peaked around $k = k_H$ and there is no correlation at large scales in the case of white noise initial condition. The suppression at small scales $k\gtrsim k_H$ is due to the DW dynamics. In fact, these features can be well explained by a simple model in which infinitely thin DWs have a uniform probability distribution on an arbitrary 1D line in space~\cite{Takahashi:2020tqv,Kitajima:2022jzz}.  In the case of inflationary fluctuations, on the other hand,
the power spectrum is nearly scale-invariant at large scales,
and suppressed at small scales as in the white noise case.
The slight tilt of the power spectrum at large scales can be understood as follows. First, the initial power spectrum is scale-invariant and grows as the scalar field rolls down to either potential minimum. Then, the enhancement factor for a mode $k$ is inversely proportional to the typical initial value of the coarse-grained field, $\bar\phi_R$, averaged over the comoving scale $R=2\pi/k$, and $\bar\phi_R$ grows as $R$ increases;
\begin{equation}
    \bar\phi_{R_1}^2- \bar\phi_{R_2}^2  \simeq 
\int^{\ln k_2}_{\ln k_1}
d\ln k' {\cal P}(k') \propto \ln \frac{k_2}{k_1} > 0,
\end{equation}
where $R_1 > R_2$, i.e. $k_1<k_2$, is assumed in the last inequality. We confirmed that this nicely explains the slightly blue-tilted spectrum at large scales.

\begin{figure}[t!]
\begin{center}  
\includegraphics[width=8cm]{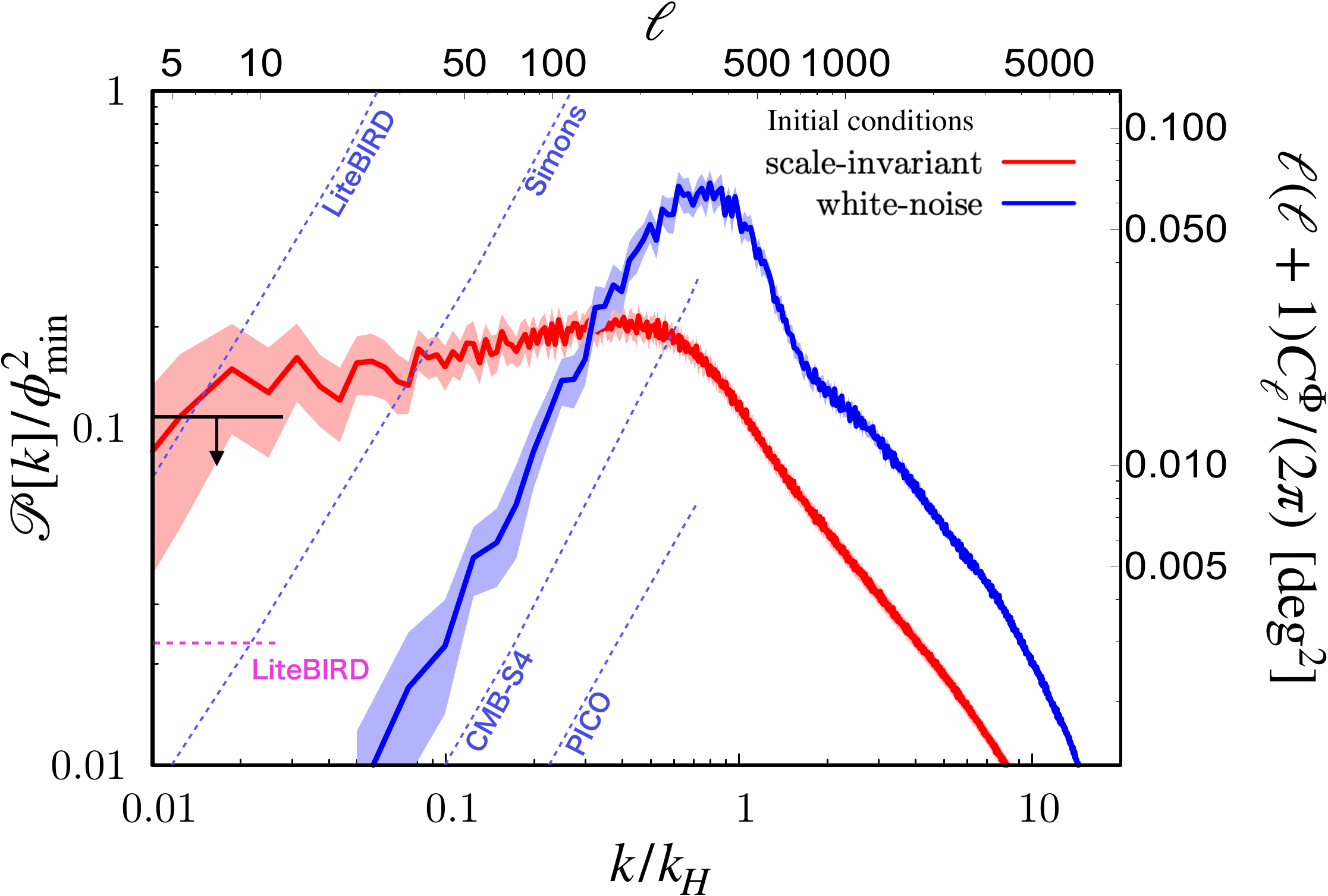}
    \end{center}
\caption{
The reduced power spectrum of the scalar field at  $\tau=10/m_0$ with the scale-invariant (red) and Gaussian white noise (blue) initial conditions in the absence of the bias, ${b_d}=0$. The degree $\ell$ and the angular power spectrum are also displayed on the top and right frame labels in the case of the anisotropic CB induced by ALP DWs (see the text).
The sensitivity reaches of various future CMB observations {for the white-noise power spectrum}
are shown by diagonal dashed lines. The horizontal solid (black) and dotted (magenta) lines  on the left edge show the current CMB bound by BICEP/Keck~\cite{BICEPKeck:2022kci} and the sensitivity reach of LiteBIRD for the scale-invariant spectrum, respectively. They are adopted from Ref.~\cite{Pogosian:2019jbt} {(see also Refs.~\cite{Matsumura:2016sri,Ade:2018sbj,Abazajian:2016yjj,Hanany:2019lle})}.} 
 \label{fig:spectrum}
\end{figure}

\textit{\textbf{An application of the DW stability to the cosmic birefringence from ALP DW.\,---\,}} 
The dynamics of the DW network is not sensitive to the detailed shape of the $Z_2$ symmetric potential, and in particular, the above argument can be applied to the sine-Gordon potential, 
\begin{equation}
V(\phi) = m_\phi^2 f_\phi^2 \left(1+\cos\left(\frac{\phi}{f_\phi}\right)\right),
\end{equation}
where $m_\phi$ and $f_\phi$ are the mass and decay constant of $\phi$. In the following $\phi$ is identified with an ALP with the above potential.
The sine-Gordon potential is the simplest of the potentials with discrete shift symmetry, 
$\f \to \f + 2\pi f_\f$,
and there are mulitple degenerate vacua. We assume that the two vacua at $\phi_{\rm min} = \pm \pi f_\phi$ are populated by the initial fluctuations, for simplicity.

The ALP is coupled to photons as
\begin{equation}
   {\cal L} \supset
  -c_\g \frac{\a_{\rm em}}{4\pi f_\f} \f F_{\mu \nu}\tl F^{\mu \nu},
\end{equation}
where $c_\gamma$ is the anomaly coefficient of ${\cal O}(1)$,
$\a_{\rm em}$  the fine-structure constant, and $F_{\mu \nu}$ and $\tl F_{\mu \nu}$ the photon field strength and its dual, respectively. Suppose that during the propagation of the CMB photon from the LSS to us, the ALP field value changes in time and space slowly enough compared to the frequency of the photon. Then the polarization plane rotates in proportion to the change in the ALP field~\cite{Carroll:1989vb,Carroll:1991zs,Harari:1992ea}.
If the ALP DWs have existed from the time of recombination to the present, the rotation angle is~\cite{Takahashi:2020tqv}
 \begin{equation}
 \label{polar}
 \D\F(\Omega) \simeq 
  0.42 {\rm~ deg } \times c_\g \left( \frac{\f_{\rm today}-\f_{\rm LSS}(\Omega) }{2\pi f_\phi}\right),
 \end{equation} 
with $\Omega$ being the spherical polar coordinates. Here $\f_{\rm today}$ and $\phi_{\rm LSS}$ is the vacuum at the solar system and  at the LSS, respectively. The parenthesis in Eq.~(\ref{polar}) is approximately equal to $1/2$ when averaged over the entire sky\footnote{
In ground-based CMB experiments, the observed rotation angle may differ by ${\cal O}(1)$because of the limited field of view, especially for ALP DWs with inflationary fluctuations.
}, and it predicts the isotropic CB, $\beta \simeq 0.21 c_\gamma {\rm \, deg}$~\cite{Takahashi:2020tqv}, independent of $m_\phi$ and $f_\phi$. This is consistent with the recent hint, $\beta=0.36\pm 0.11 {\rm\, deg}$~\cite{Minami:2020odp,Diego-Palazuelos:2022dsq}.

Unlike models based on spatially uniform ALP~\cite{Minami:2020odp, Fujita:2020ecn, Mehta:2021pwf,Choi:2021aze, Nakagawa:2021nme, Gasparotto:2022uqo, Murai:2022zur,Lin:2022khg}, the ALP DWs predict anisotropic CB that reflects the DW distribution on the LSS.
We 
can map the reduced power spectrum obtained above to the LSS by using the relations,
$
\sqrt{\ell (\ell+1)}/d_L \leftrightarrow k, \ell (\ell+1) C_\ell^\F/2\pi\beta^2 \leftrightarrow  {\cal P}[k]/\phi_{\rm min}^2$, where $d_L$ is the comoving distance to the LSS, and $C_\ell^\F$ is the angular power spectrum of $\Delta \Phi(\Omega)$~\cite{Bernardeau:2010ac,Kitajima:2022jzz}. 
The mapped $\ell$ and  $C_\ell^\F$ are shown on the  upper and right axes of the panel in Fig.~\ref{fig:spectrum}, where $\beta=0.36\, {\rm deg}$ is assumed. Note that  the anisotropic CB is correlated with the isotropic one as $C_\ell^\F \propto \beta^2$. 
 {We also show the future reaches and present bound for the {white noise and scale-invariant power spectra}, including the BICEP/Keck bound on
 the scale-invariant spectrum, $\ell(\ell+1) C_\ell^\F/(2\pi) < 0.014\, {\rm deg}^2$ (95 \% CL)~\cite{BICEPKeck:2022kci}.}
While the prediction shown in the figure barely satisfies the current limit, one can relax the tension by taking slightly smaller $\beta$ or including the effects of cosmic variance and/or bias, while explaining the isotropic CB\cite{Minami:2020odp, Diego-Palazuelos:2022dsq}. We have checked that, even for $b_d= 1$, the reduced power spectrum  gets suppressed up to by
$20\%$ at large scales and is stable over time.
Note also that the lowest multipoles $\ell < 20$ are most important for constraining the scale-invariant CB.
 Thus, the predicted anisotropic CB can be confirmed or refuted in near-future experiments including LiteBIRD.

\textit{\textbf{Conclusions.--}} 
{We have shown that DW network is very stable against the bias of initial inflationary distribution. This is at odds with the understanding of the past few tens of years. The unexpected stability is due to correlations on large scales generaterd by inflation, which lead to the formation of DWs of superhorizon size. Once formed, such DWs are very stable due to  causality in the present set-up. Thus, various studies on the production of the gravitational waves, baryon asymmetry, dark matter, and CB from DWs must be reconsidered. We studied the last one in this Letter, but there is still room for further research on the rest.}

\section*{Acknowledgments}
This work is supported by the MEXT scholarship (grant number: 203069) (D.G), JSPS Core-to-Core Program (grant number: JPJSCCA20200002) (F.T.),  JSPS KAKENHI Grant Numbers  19H01894 (N.K.), 20H01894 (F.T. and  N.K.), 20H05851 (F.T.,  N.K. and W.Y.), 21H01078 (N.K.), 21K20364 (W.Y.),  22K14029 (W.Y.), and 22H01215 (W.Y.).

\clearpage

\appendix
\title{Supplemental Material}

\subsection{Time evolution of the DW network}

In the main text we have considered the characterization of the DW network through the total DW length per Hubble horizon, $L_{\mathrm{DW}}/H^{-1}$. There are other interesting measures in the system, like the total positive or negative area.  We define the ratio $r$ as
\begin{equation}
\label{eq:ratio}
    r = \frac{2 S_{-}}{S_{+}+S_{-}},
\end{equation}
where $S_{+}(S_{-})$ is the total area where field values of $\phi$ are positive (negative). Here $r$ is defined so that it is bounded as $r\in[0,2]$, making it easy to study in a logarithmic plot.  It is straightforward to define the ratio for the 3D case. 
In a $d$ dimensional space, $L_{\mathrm{DW}}/H^{-(d-1)}$ and $r$ encode information on the $d-1$ dimensional hypersurface and $d$ dimensional volume, respectively. In principle, these are independent measures, all their correlations stemming from the DW dynamics. 

We show in Fig.~\ref{fig:ratio} the time evolution of the ratio $r$ 
for the inflationary fluctuations (top) and Gaussian white noise (bottom). As with the DW length, the ratio decreases only slowly even for large biases in the case of inflationary fluctuations.
Also, as we can see in Fig.~\ref{fig:ratio}, after $m_0\tau=5$ when the DW network is fully formed, {the evolution of $r$ can be well described by an exponential decay over the time range shown. However, unless $b_d$ is very large, the DW lifetime in the case of inflationary fluctuations is so long that it is not possible to determine numerically whether $r$ eventually approaches $0$ and the DW network decays completely, or whether $r$ asymptotes to a certain value. }

\begin{figure}[t!]
\includegraphics[width=8cm]{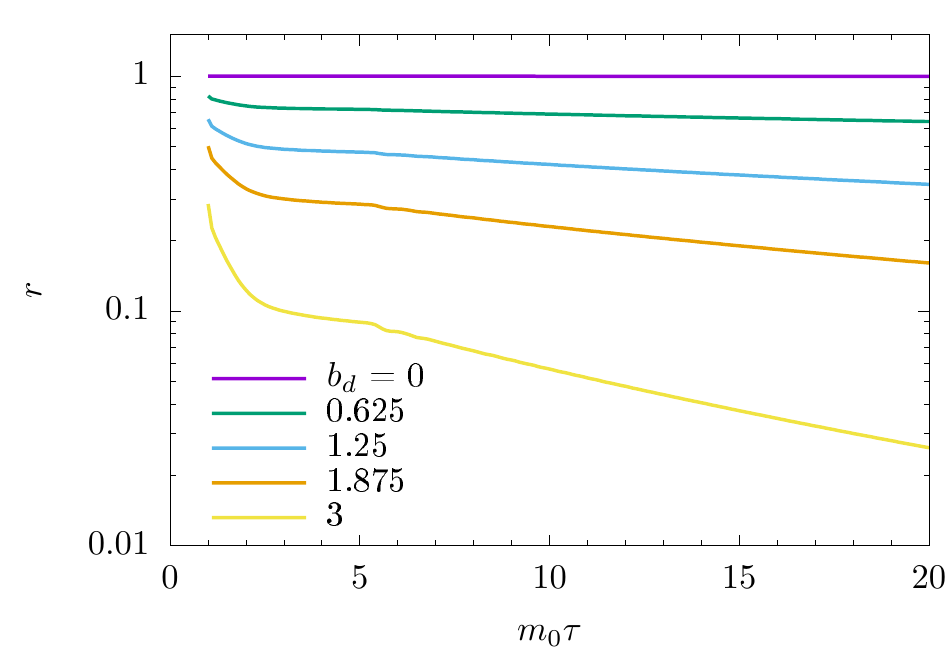}
\includegraphics[width=8cm]{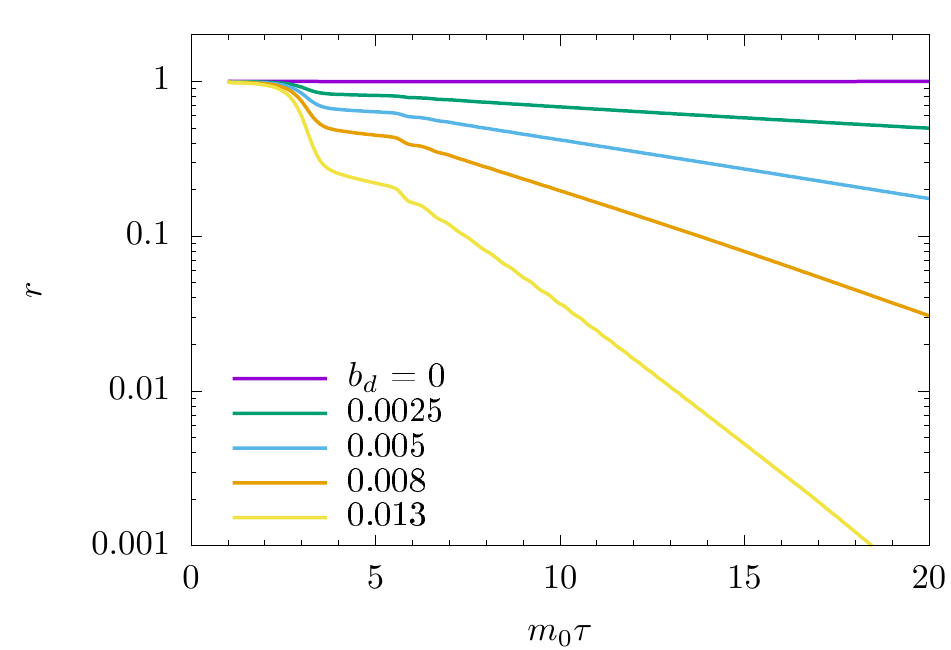}
\caption{The ratio $r$ as defined in Eq.~\eqref{eq:ratio} as a function of $\tau$, based on 500 simulations. Initial fluctuations are set with inflationary fluctuations (top) and Gaussian white noise (bottom). The top line in each panel corresponds to $b_d=0$ and the lower lines are for progressively more biased initial fluctuations.}
\label{fig:ratio}
\end{figure}

In Fig.~\ref{fig:snapshots} we show snapshots of the DW network for various biases in the case of inflationary fluctuations. For $b_d=0.15$ —a very large bias in the case of white noise DW networks— we can see that even at later times there are still ``infinitely" large DWs within the box. Once the bias becomes large enough, as is in the case of $b_d=1.25$, then no ``infinitely" large DWs are found at later times. {This suggests that for the inflationary  fluctuations, if the fluctuation at the largest scale of interest is greater than the bias,  DWs separating domains of the largest size are likely formed. Such DWs will be
stable or very long-lived due to the causality. 
Thus, there might be a critical value of the bias parameter $b_{d,\mathrm{crit}}$, above which the DWs are finite and completely decay in a finite time.
On the other hand, for biases smaller than the critical value the DW length per horizon as well as the ratio $r$ may asymptote to some constant values. 
This kind of arguments will help us to understand the stability of the DW network originated from scale-dependent superhorizon fluctuations. For instance, if the initial fluctuations are not scale-invariant, but blue-tilted, there may be an upper bound on the size of the DWs, which could be many orders of magnitudes larger than the Hubble horizon at the formation time. In this case, the largest DW size roughly determines the lifetime of the whole DW network. 
}

\begin{figure}[t!]
    \centering
\begin{tabular}{ c c }
 \includegraphics[width=4cm]{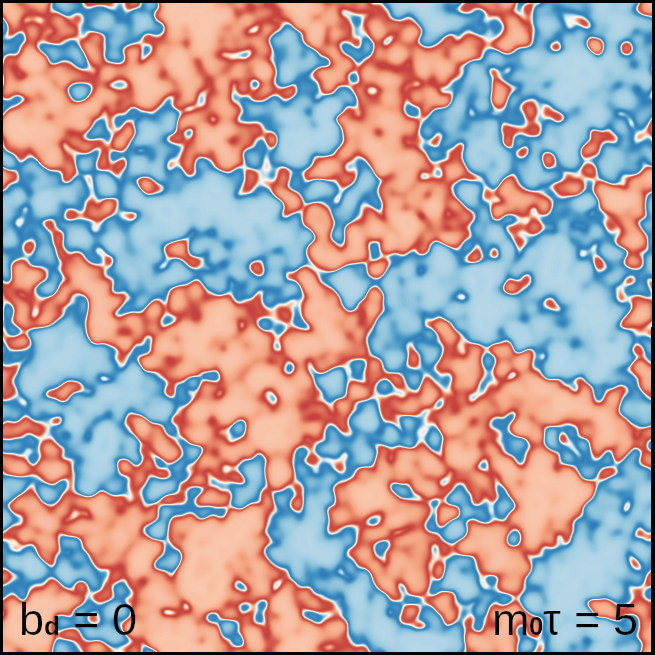} &  \includegraphics[width=4cm]{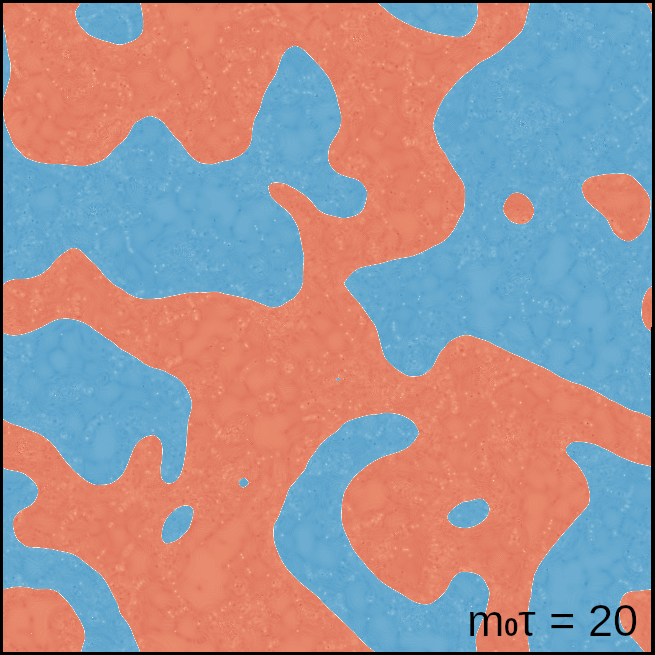} \\ 
  \includegraphics[width=4cm]{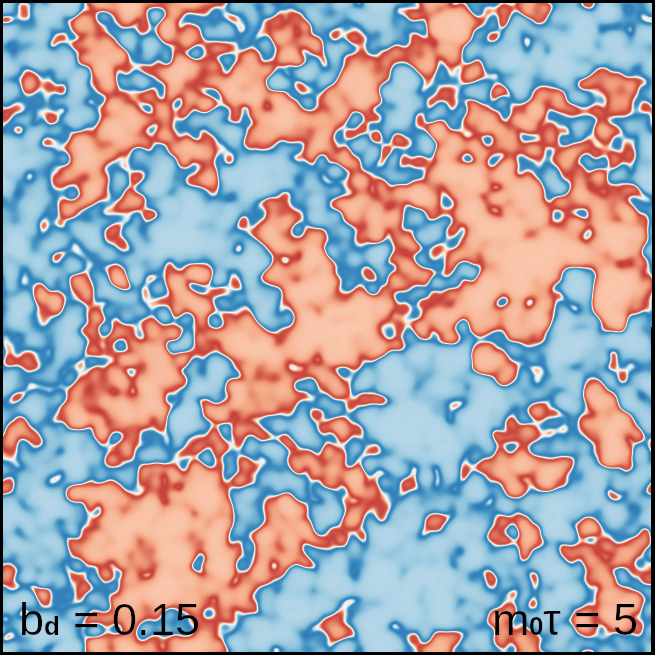} &  \includegraphics[width=4cm]{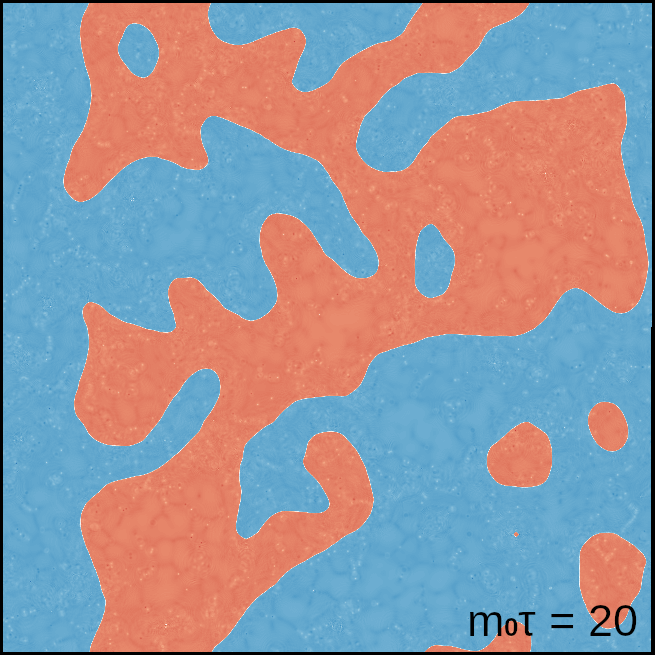} \\  
  \includegraphics[width=4cm]{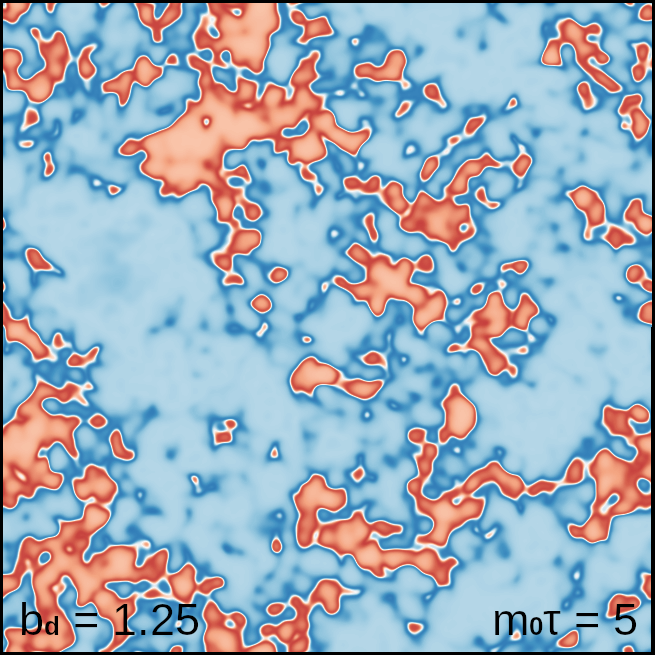} &  \includegraphics[width=4cm]{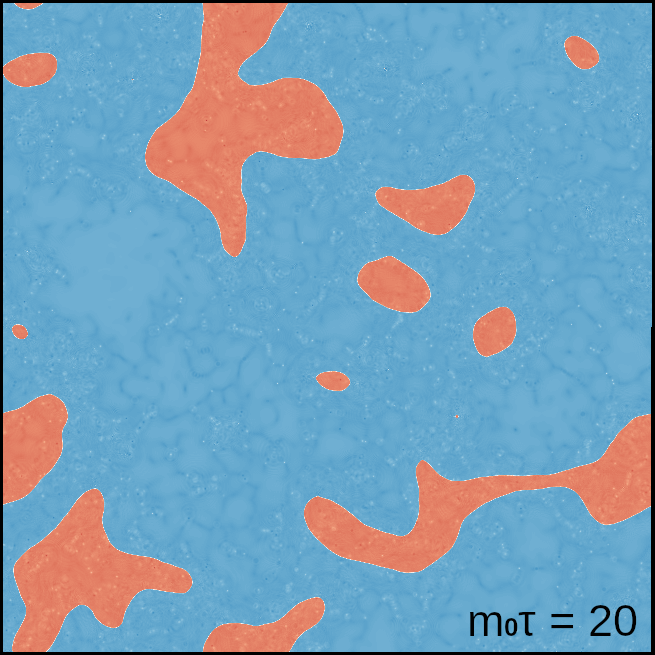} \\  
\end{tabular}
\caption{Snapshots of the DW network for $b_d=0$ (top row), $0.15$ (middle row) and $1.25$ (bottom row), for times $m_0\tau=5$ (left column) and $m_0\tau=20$ (right column). Blue and yellow regions correspond to $\phi > 0$ and $\phi < 0$ respectively. Initial fluctuations are set with inflationary fluctuations. The box contains { $\sim 40^2 (10^2) $  horizons at $m_0 \tau = 5 (20)$.}}
\label{fig:snapshots}   
\end{figure}

\subsection{A DW model of subhorizon fluctuation with bias}
\label{asym}
In the main text, we focused on the inflationary fluctuations, while the Gaussian white noise is considered only for comparison. On the other hand, the formation and evolution of DWs based on such biased white noise initial fluctuations have been investigated in the past literature~\cite{Lalak:1993bp,Lalak:1994qt,Coulson:1995nv,Larsson:1996sp,Correia:2014kqa,Correia:2018tty}. Therefore, it would be useful to provide an explicit UV model that actually realizes such initial condition.

One way is to introduce another axion, $a'$, with a mixing with an axion $a$ that constitutes the DW in the low energy. If the global U(1) symmetry for $a'$ is spontaneously broken after inflation, axionic strings are produced. On the other hand, the U(1) symmetry for $a$ is assumed to be already spontaneously broken during inflation. If a combinations of two U(1)'s for $a$ and $a'$ is explicitly broken, those strings annihilate when the corresponding DWs are generated. Then, we are left with almost white noise initial condition for the remaining lighter degrees of freedom~\cite{Takahashi:2020tqv}.  
 
As another example, let us build a model with  approximate $Z_2$ symmetry, such that  the adjacent vacua are physically identical and therefore strictly degenerate in energy. 
We introduce an axion, $a$, which enjoys the discrete shift symmetry, with an approximate (dark) charge conjugation symmetry, $C_{\rm dark}$. 
The interacting 
Lagrangian is given as, 
\beq 
{\cal L}=-y\Phi \bar{\Psi}_L \Psi_R- M \bar{\Psi}_L \Psi_R+ {\rm h.c.}
\eeq 
Here $\F$ is a Peccei-Quinn (PQ) field for the axion, and $\P_{L,R}$ are PQ fermions, while the fermion mass $M$ is the explicit breaking of the PQ symmetry. 
Under the $C_{\rm dark}$ symmetry, the fields transform as $\bar{\Psi}_L\leftrightarrow \Psi_R, \Phi\leftrightarrow \Phi^*$, and therefore, both $y$ and $M$ must be real.

Now let us consider the low energy effective theory for the axion, $a$, after the PQ spontaneous symmetry breaking.
The Nambu-Goldstone boson, $a$, resides in the phase as $\Phi= \frac{f_a}{{\sqrt{2}}} \exp{[i a/f_a]}$, and it obtains the potential of 
\beq 
{\cal L}_{\rm eff}=-y \frac{f_a}{\sqrt{2}} e^{i a/f_a} \bar{\Psi}_L \Psi_R- M \bar{\Psi}_L \Psi_R+ {\rm h.c.}. 
\eeq 
The axion effective (thermal) potenital is given by
\beq
\d V_{\rm eff} = -\frac{1}{64\pi^2}M_{\rm eff}^4 \(\log \frac{M_{\rm eff}^2}{\mu^2}-\frac{3}{2}\)- \frac{T^4}{\pi^2}  J(M_{\rm eff}^2/T^2)
\eeq
with 
\beq 
J(y^2)= \int_0^{\infty}{dx x^2 \log(1+  e^{-\sqrt{x^2+y^2}})}
\eeq
and $M_{\rm eff}^2\equiv \ab{\frac{y f_a}{\sqrt{2}}e^{ia/f_a}+ M}^2\supset \sqrt{2}(y f_a M) \cos(a/f_a).$ 
Here we assume that $\Psi$ is  thermalized at temperature $T$. 
We obtain a thermal mass contribution at $T^2\gg M_{\rm eff}^2$ as 
\beq
\delta V_{\rm eff}\supset -\frac{T^2}{24} M_{\rm eff}^2\supset  -\frac{T^2}{24} \sqrt{2} y f_a M \cos(a/f_a).
\eeq 
Since both $y f_a $ and $M$ are real thanks to the $C_{\rm dark }$ symmetry, 
the thermal mass at high enough $T$ can bring the axion to $a=0$ if $y M>0.$ 

We may also have another source for the axion potential from e.g. non-perturbative effects of non-Abelian gauge symmetry, in which case the total potential is
\beq V(a)= \Lambda^4 \cos(a/f_a) +\delta V_{\rm eff}.
\eeq 
As long as the explicit breaking term satisfies $C_{\rm dark},$ the potential always has the symmetric enhanced point at $a=0$ since $a\to -a $ under $C_{\rm dark}.$ 
Thus $a=0$ is either the potential maximum or minimum. 
Since the axion is at the hilltop due to the thermal mass at a high enough temperature,  
the initial condition for the domain wall formation is set by thermal effects. 
At the low temperature, the symmetry enhancement point may turn out to be a local maximum, and thus the domain wall is formed at $H\sim m_a$ with $m_a$ being the axion mass in the vacuum. 

If we introduce a tiny breaking the $C_{\rm dark}$ symmetry, e.g., by adding a phase, $\theta\neq 0$ for the coupling, $y\to \exp{(i \theta)} y$. Then the thermal potential, in general, does not necessarily bring the axion exactly at the origin, and the misalignment will induce the bias of the axion distribution.
On the other hand, in spite of such bias, we still have degenerate vacua due to the periodicity of the
potential, i.e., the potential does not change with $a\to a+2\pi f_a$.

\bibliography{ref}
\end{document}